\documentclass[sigconf]{acmart}
\usepackage{hyperref}
\usepackage{url}
\usepackage{graphicx}
\usepackage{amsmath}
\usepackage{mathtools}
\usepackage{amsthm}
\usepackage{multirow}
\usepackage[lined,ruled]{algorithm2e}
\usepackage{enumitem}
\usepackage{graphicx}
\usepackage[utf8]{inputenc} 
\usepackage[T1]{fontenc}    
\usepackage{hyperref}       
\usepackage{url}            
\usepackage{booktabs}       
\usepackage{amsfonts}       
\usepackage{nicefrac}       
\usepackage{microtype}      
\usepackage{xcolor}         
\usepackage[nomargin,inline,marginclue,draft]{fixme}
\usepackage{enumitem}
\usepackage{subfig}
\usepackage{multirow}
\AtBeginDocument{%
  }

\renewcommand\footnotetextcopyrightpermission[1]{}
\begin{document}

\title{A Large-scale Dataset with Behavior, Attributes, and Content of Mobile Short-video Platform}


\author{Yu Shang}
\affiliation{%
  \institution{Department of Electronic Engineering\\ Tsinghua University}
  \city{Beijing}
  \country{China}}
\email{shangy21@mails.tsinghua.edu.cn}

\author{Chen Gao}
\affiliation{%
  \institution{BNRist\\ Tsinghua University}
  \city{Beijing}
  \country{China}}
\email{chgao96@tsinghua.edu.cn}

\author{Nian Li}
\affiliation{%
  \institution{Department of Electronic Engineering\\ Tsinghua University}
  \city{Beijing}
  \country{China}}
\email{linian21@mails.tsinghua.edu.cn}

\author{Yong Li}
\affiliation{%
  \institution{Department of Electronic Engineering\\ Tsinghua University}
  \city{Beijing}
  \country{China}}
\email{liyong07@tsinghua.edu.cn}


\begin{abstract}
   Short-video platforms show an increasing impact on people’s daily lives nowadays, with billions of active users spending plenty of time each day. The interactions between users and online platforms give rise to many scientific problems across computational social science and artificial intelligence. However, despite the rapid development of short-video platforms, currently there are serious shortcomings in existing relevant datasets on three aspects: inadequate user-video feedback, limited user attributes and lack of video content. To address these problems, we provide a large-scale dataset with rich user behavior, attributes and video content from a real mobile short-video platform. This dataset covers 10,000 voluntary users and 153,561 videos, and we conduct four-fold technical validations of the dataset. First, we verify the richness of the behavior and attribute data. Second, we confirm the representing ability of the content features. 
   Third, we provide benchmarking results on recommendation algorithms with our dataset. Finally, we explore the filter bubble phenomenon on the platform using the dataset.
   We believe the dataset could support the broad research community, including but not limited to user modeling, social science, human behavior understanding, etc. 
   The dataset and code is available at \href{https://github.com/tsinghua-fib-lab/ShortVideo_dataset}{https://github.com/tsinghua-fib-lab/ShortVideo\_dataset}.
\end{abstract}





\maketitle

\section{Introduction}
Short-video platforms such as Tiktok are redefining how users access information on the web. 
With the widely deployed algorithms of personalized recommendation, the platform can infer user preferences from users' behavioral logs and users' demographics, which helps alleviate information overload.
However, the intelligence algorithm has also brought critical social concerns including echo chamber~\citep{ge2020understanding}, filter bubble~\citep{nguyen2014exploring}, user addiction~\citep{montag2018internet}, etc. 

Despite the fast growth of short-video platforms, there is no publicly available dataset that can well support the research of improving user modeling and alleviating the negative impact caused by AI algorithms.
The existing public datasets of short-video platforms are released by researchers and engineers of recommendation algorithms. However, they only focus on user feedback while are quite limited in supporting the research areas beyond recommendation algorithms.
In specific, 
KuaiRec~\citep{gao2022kuairec} is a fully observed micro-video dataset with additional user and video attributes collected from Kuaishou platform. 
REASONER~\citep{chen2023reasoner} is an explainable recommendation dataset, containing some basic attributes and ground truths for multiple explanation purposes. 
Tenrec~\citep{yuantenrec} is a large-scale multipurpose benchmark dataset for recommendation tasks, which also contains interactions between users and videos.
MicroLens~\citep{ni2023content} is a recently released micro-video dataset with various modality content of videos.
Although these datasets may support the research of machine learning models for personalized user modeling or recommendation. 
They are suffering from limitations including (1) lack of video content (2) inadequate user-video feedback and (3) the shortage of user attributes.

\begin{figure*}[t!]
\centering
\subfloat[User interface and behaviors on the platform.]{\includegraphics[width=0.5\linewidth]{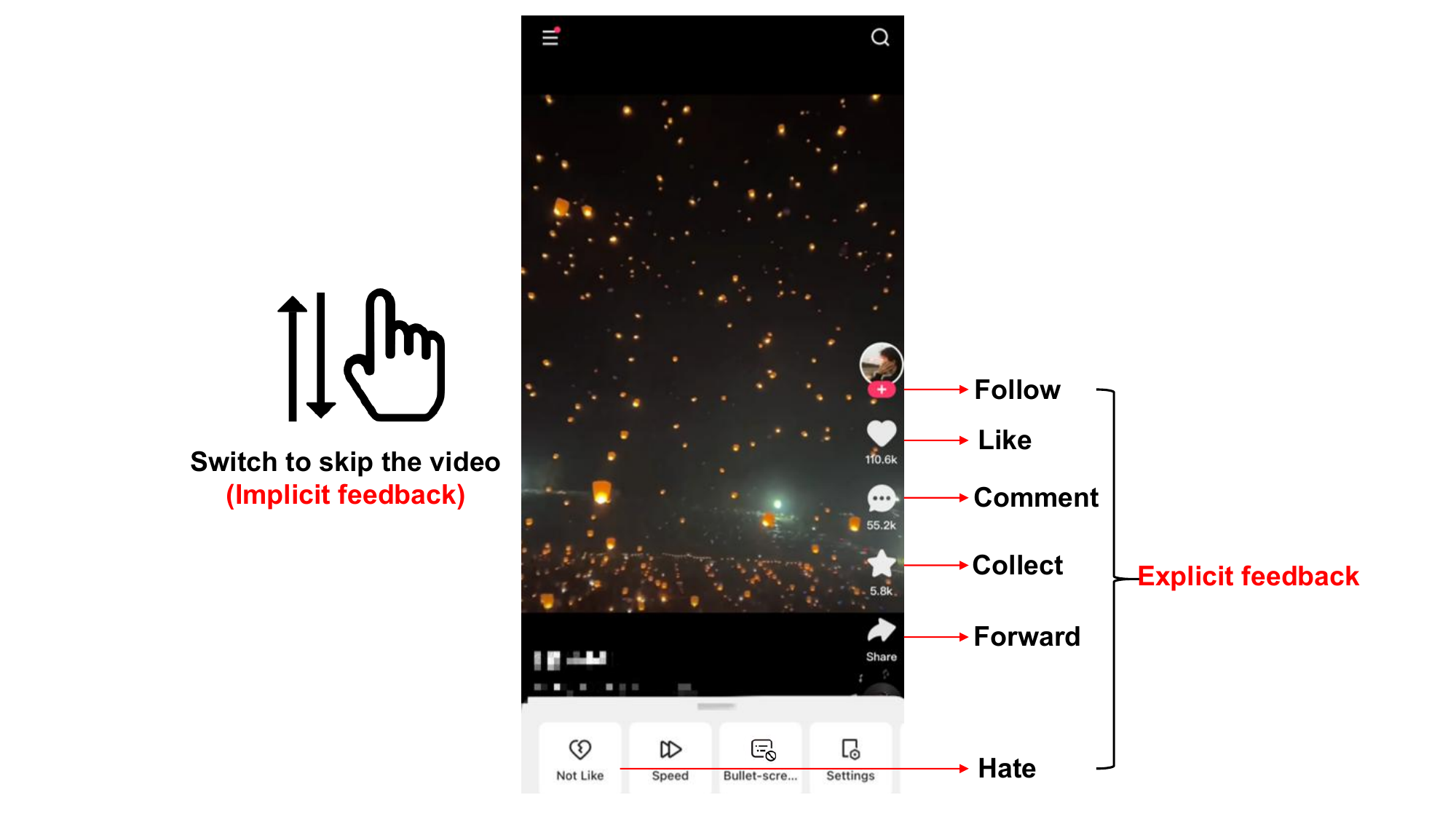}\label{fig:interface}}
\subfloat[Dataset Overview.]
{\includegraphics[width=0.5\linewidth]{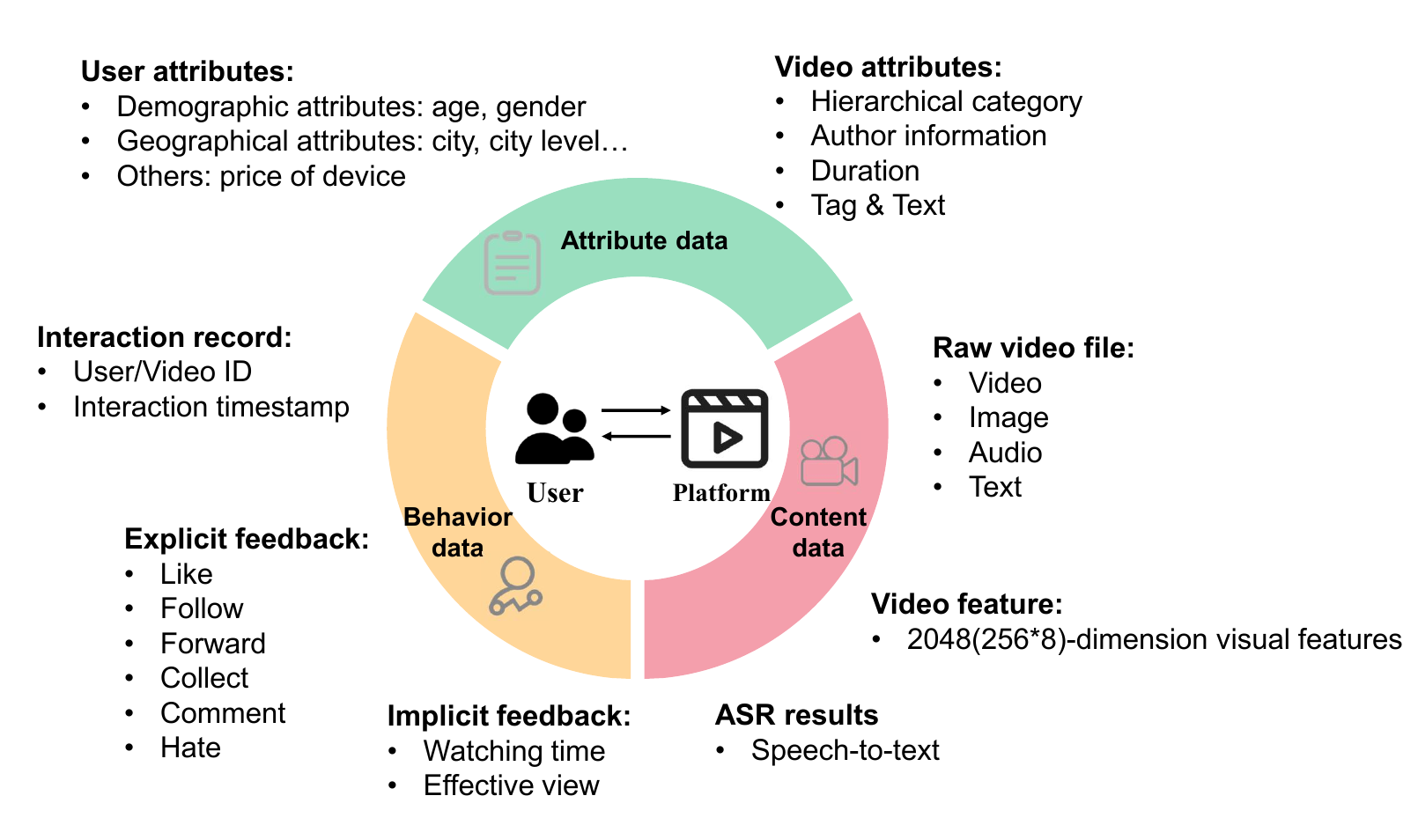}\label{fig:data_overview}
}
\caption{The illustration of user interface and behaviors on the platform (a) and an overview of the dataset (b).}
\label{fig:data_overview}
\end{figure*}

In this paper, we provide a large-scale short-video dataset from 10,000 volunteers on one of the largest short-video platforms. The data collection procedure strictly follows privacy and ethical regulations. 
In terms of data scale, our dataset covers 1,019,568 interactions between 10,000 users and 153,561 videos, which is larger than most existing released datasets, e.g., Kuairec~\citep{gao2022kuairec} and REASONER~\citep{chen2023reasoner}.
Besides, compared with datasets used in real industrial scenarios, our dataset has a similar scale to the datasets used to conduct offline experiments and analysis, such as the dataset of WeChat Channel~\footnote{https://algo.weixin.qq.com/2021/problem-description} and Tiktok~\footnote{https://www.biendata.xyz/competition/icmechallenge2019}. 
Therefore, our dataset can satisfy the requirements of both academic and industrial scenarios.
Our dataset outperforms existing public datasets in three aspects: user behavior, user/video attributes, and video content, which is detailed in Section 2. 
We further conduct a four-fold validation of the dataset usability in Section 3, including data distribution analysis, video content quality validation, benchmarking recommendation algorithms and the filter bubble phenomenon study.
Our key contributions are as follows:
\begin{itemize}[leftmargin=*]
    \item We collect a large-scale dataset from a real mobile short-video platform, covering rich user behavior, attributes and video content, which are scarce in existing datasets.
    \item We conduct comprehensive technical validation of the wide coverage of behavior and attribute data and the quality of content data in the dataset.
    \item We provide a sufficient exploration of the potential research directions with our dataset, covering broad research communities such as user modeling and studying filter bubbles.
\end{itemize}


\section{Data Description}
We collected six months of interaction data from 10,000 hired volunteers (excluding users under 20) with their consent, in the paper we focus on analyzing the first week’s data for a quick release. 
An overview of the dataset is shown in Figure \ref{fig:data_overview}. 


\subsection{Behavior Data}
We asked the volunteers to install a proxy agent on their mobile devices to record the interactions, which was fully acknowledged by the volunteers.
For each interaction record, we record the basic information including anonymous user/video ID and timestamp of the interaction. More importantly, we have collected rich user behavior which could convey diverse preference signals of users including explicit feedback (like, comment, follow, forward, collect and hate behavior) and implicit feedback (watching time).  

\par

\subsection{Attribute Data}
Compared with existing datasets, the dataset contains richer attribute information about the users and videos. Specifically, there are 9 video-side attributes and 6 user-side attributes in the dataset.

For the video side, we have fully exploited the key attributes of micro-videos including video category, author information, duration, video text and video tag. 
\begin{itemize}[leftmargin=*]
\item \textbf{Video category}. 
We establish the hierarchical categories by conducting K-means clustering of the video title and manually labeling their categories, which can cover most of the collected videos. 
The category of videos is divided into 3 levels hierarchically (Category \uppercase\expandafter{\romannumeral1}, Category \uppercase\expandafter{\romannumeral2}, Category \uppercase\expandafter{\romannumeral3}). 
For example, for a video recording a \textit{hockey game}, Category \uppercase\expandafter{\romannumeral1} is ``\textit{sports}'', Category \uppercase\expandafter{\romannumeral2} is ``\textit{ball game}'' and Category \uppercase\expandafter{\romannumeral3} is ``\textit{hockey}''. Statistically there are 37 kinds of Category \uppercase\expandafter{\romannumeral1}, 281 kinds of Category \uppercase\expandafter{\romannumeral2} and 382 kinds of Category \uppercase\expandafter{\romannumeral3}. 
The hierarchical category organization possesses richer semantic information compared with existing datasets, which could support multi-level video content understanding. 

\item \textbf{Author information}. 
The dataset includes 81,870 video authors, with collected information such as author ID and number of fans.
\item \textbf{Duration}. The dataset includes videos of varying durations, ranging from under 30 seconds to over 5 minutes, with most being relatively short (less than one minute). 

\item \textbf{Video title and tags}. Our dataset includes 140,341 unique titles across all videos. Video tag is added manually by authors shown in each video for brief video descriptions, such as "\textit{delicious food recommendation}" and "\textit{travel tips}". 
 There are a total of 79,705 unique tags in all collected videos. 
\end{itemize}

For the user side data, the dataset records comprehensive user attributes including demographical, and geographical characteristics, which are not adequate in existing datasets. Demographic and geographic data were coarse-grained, such as city-level location, ensuring that users could not be identified.
\begin{itemize}[leftmargin=*]
\item \textbf{Demographic attributes}. The demographic attributes include gender and age of users, which are basic for the social-concerned research on short-video platforms like user addiction. 
In the data analysis of the paper, we do not exclude the minor users but their data has been removed in the actual dataset.
\item \textbf{Geographical attributes}. The geographical attributes contain the city each user lives in, the city's level (from first-tier to fifth-tier) and the community type (country, urban area and town). The geographical characteristics could complement the user-side profile and support the analysis of regional user differences.
\item \textbf{Other attributes}. 
Our dataset records the price of user devices, which is seldom covered by previous datasets.  Considering privacy protection, we only collect the phone's model name from volunteers and then retrieve the corresponding phone price.
\end{itemize}

\subsection{Content Data}
The video content is the most novel component of our dataset, which has always been neglected by existing datasets. Our dataset includes the raw files of 153,561 videos watched by users, totaling 3,998 hours in duration and 3.2 TB in size.
We preprocess the raw videos for easier use. Each video is divided into 8 equal-length clips, and for each clip, we extract a 256-dimensional visual feature using both
pre-trained ResNet and ViT. Additionally, we provide bilingual (Chinese and English) ASR text generated with SenseVoice-Small~\citep{speechteam2024funaudiollm} and LLaMA3-8B for translation.

\section{Technical Validation and Applications}
To ensure the technical validity of the data, we validate it from four aspects. First, we assess the richness and diversity of the data, including interaction distribution and attribute diversity. Second, we validate the quality of the content data, particularly the preprocessed video features, using clustering visualization. Third, we establish a benchmark for recommendation algorithms. Lastly, we explore the filter bubble phenomenon within our dataset.

\subsection{Data Richness Validation}
\begin{figure}[!t]
\centering
\subfloat[User-side interaction distribution.]{\includegraphics[width=0.5\columnwidth]{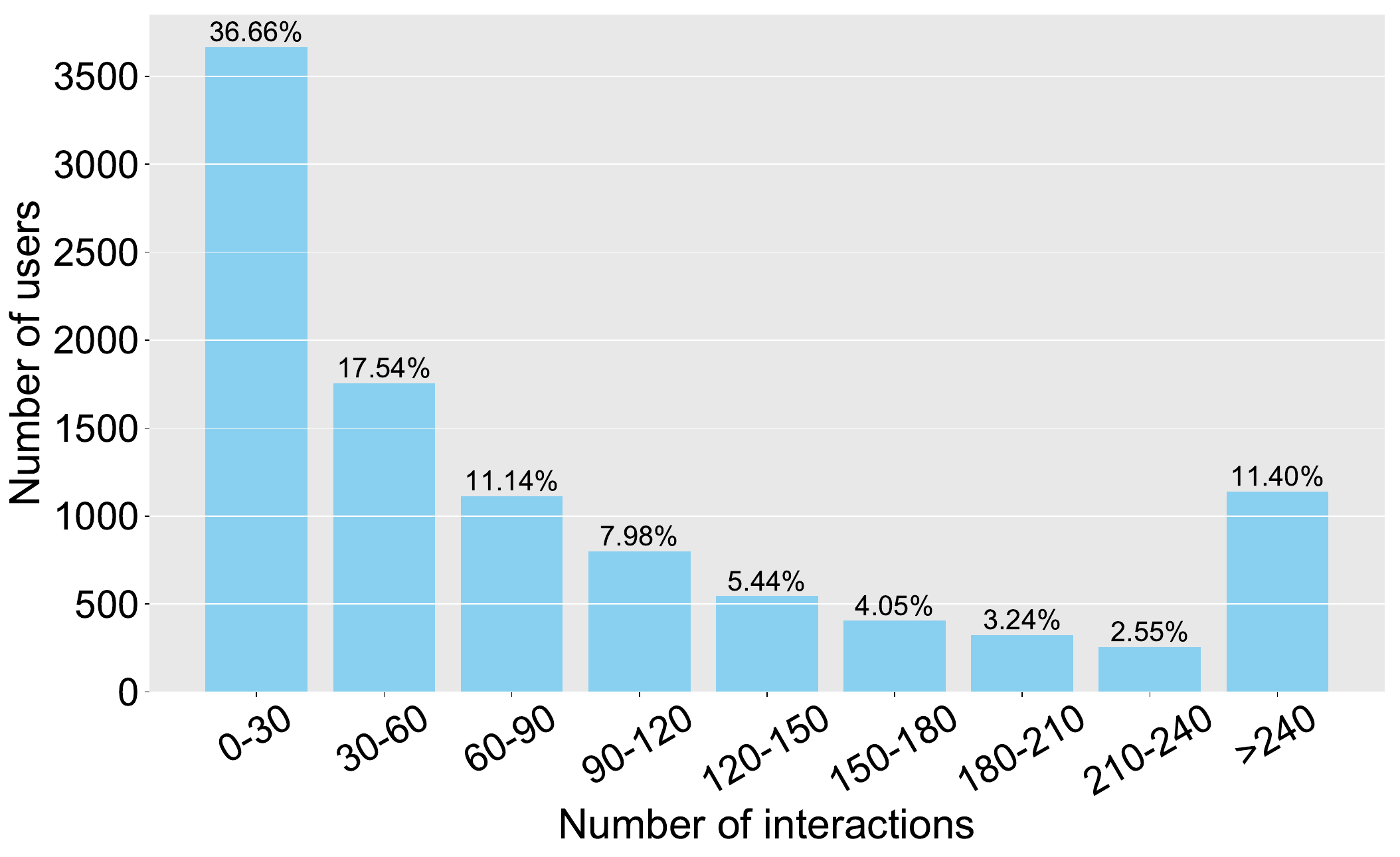}\label{fig:u_inter}}
\subfloat[Video-side interaction distribution.]{\includegraphics[width=0.5\columnwidth]{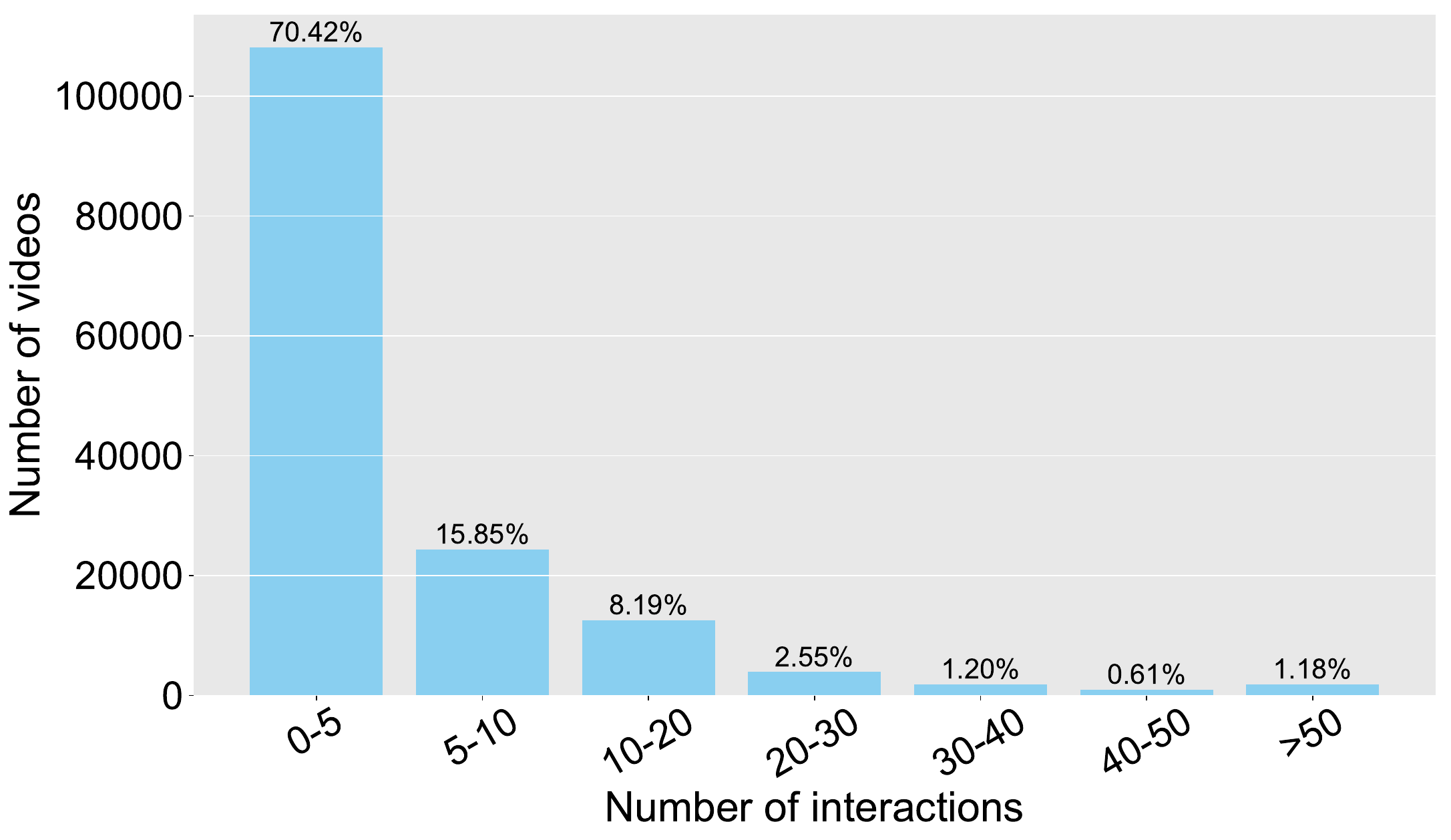}\label{fig:v_inter}}
\caption{Interaction number distribution of (a) users and (b) videos.}
\vspace{-3mm}
\label{fig:inter}
\end{figure}




\subsubsection{Interaction Distribution Analysis} 
We categorize users and videos based on their interaction frequency, as shown in Figure \ref{fig:inter}, illustrating a wide range of user activity and video exposure. As depicted in Figure \ref{fig:u_inter}, the number of users decreases with higher interaction frequency, reflecting the fact that highly active users are a minority. For videos, Figure \ref{fig:v_inter} shows varying interaction counts, ranging from 1 to 1,193. In summary, the dataset encompasses a diverse set of users and videos in terms of interaction frequency.
\begin{figure}[t]
\centering
\subfloat[User gender distribution.]
{\centering\includegraphics[width=0.48\linewidth]{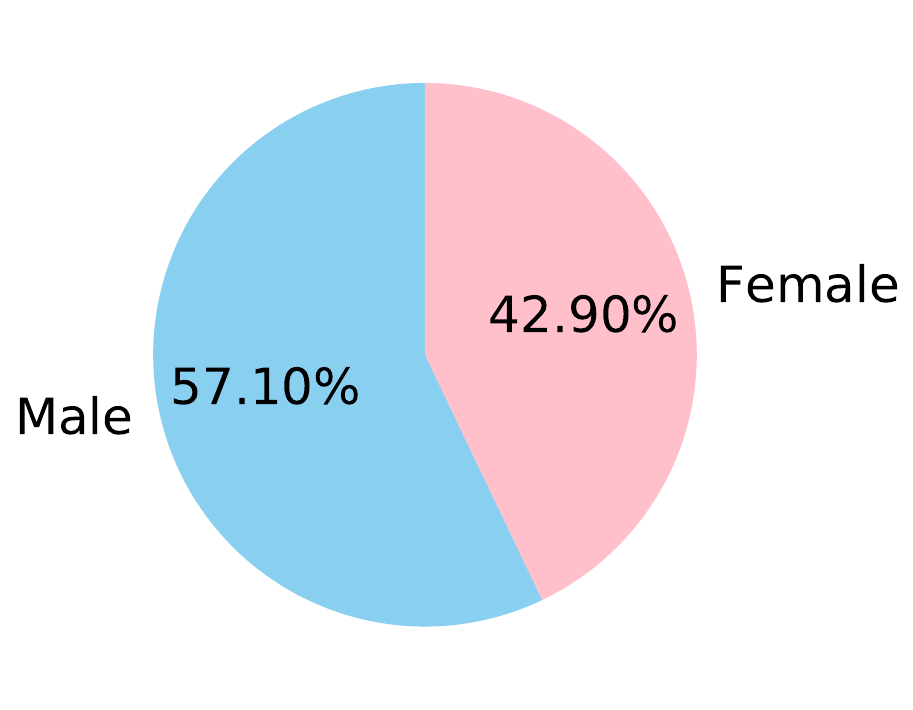}
\label{fig:gender}}
\subfloat[User City-level distribution.]{
\includegraphics[width=0.48\linewidth]{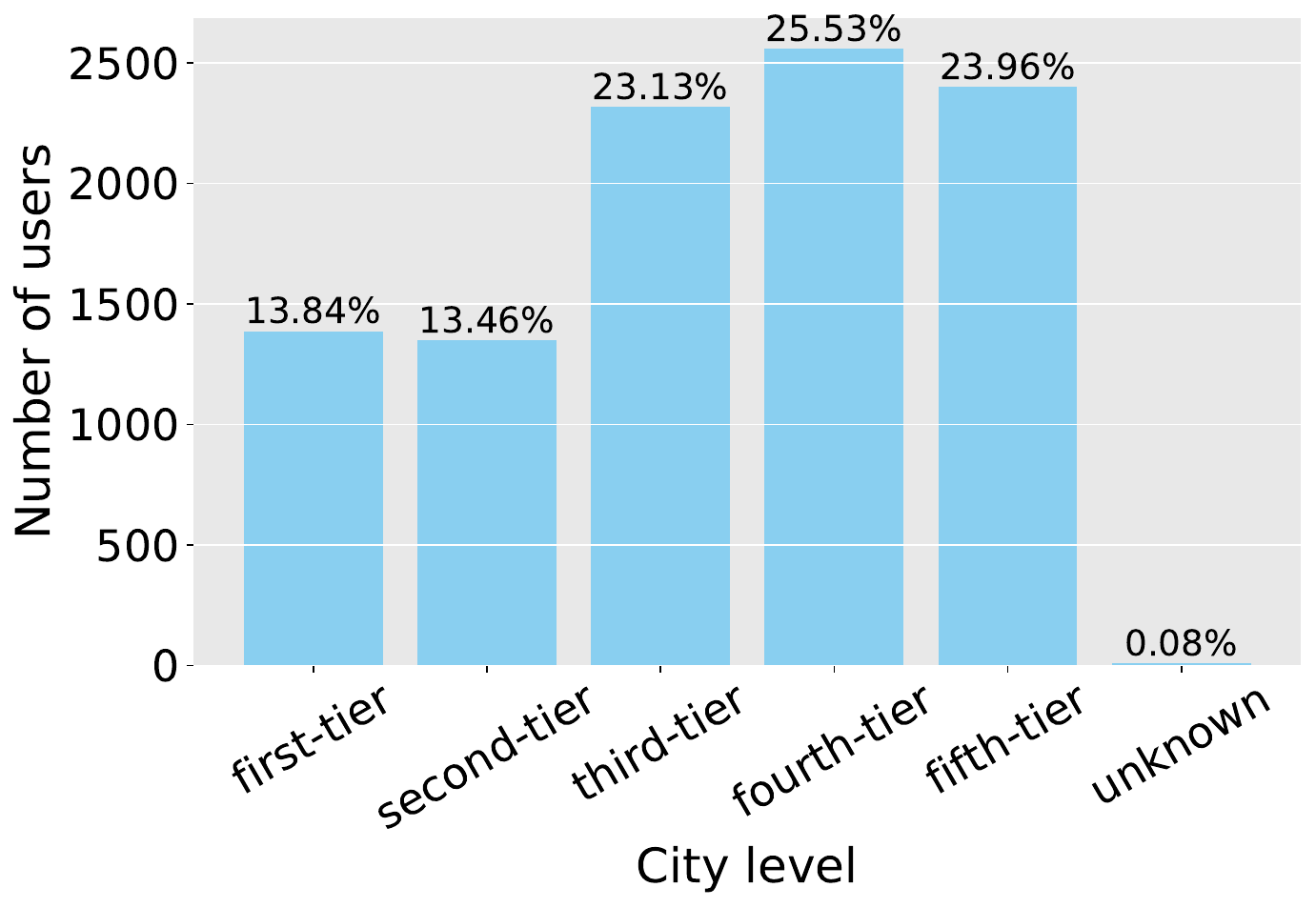} 
\label{fig:city-level}}\hfill

\caption{Distribution of some key fields in user attributes.} 
\label{fig:u_attr}
\vspace{-5mm}
\end{figure}

\subsubsection{User Attribute Coverage Analysis}
In order to validate the diversity of users, we have analyzed the distribution of some key attributes including demographic data (gender) and geographical data (city level) as shown in Figure~\ref{fig:u_attr}. Among the volunteers, 57.1\% are male and 42.9\% female, reflecting a nearly balanced distribution. This aligns closely with the official platform report of 56\% male and 44\% female users, validating the reliability of the dataset. Geographically, the dataset covers diverse users from 373 cities and 5 city levels (first to fifth tier).

\subsection{Quality Validation of Video Content Data}
\begin{figure}[t!]
\centering
\subfloat[Category \uppercase\expandafter{\romannumeral1}]
{\includegraphics[width=0.5\linewidth]{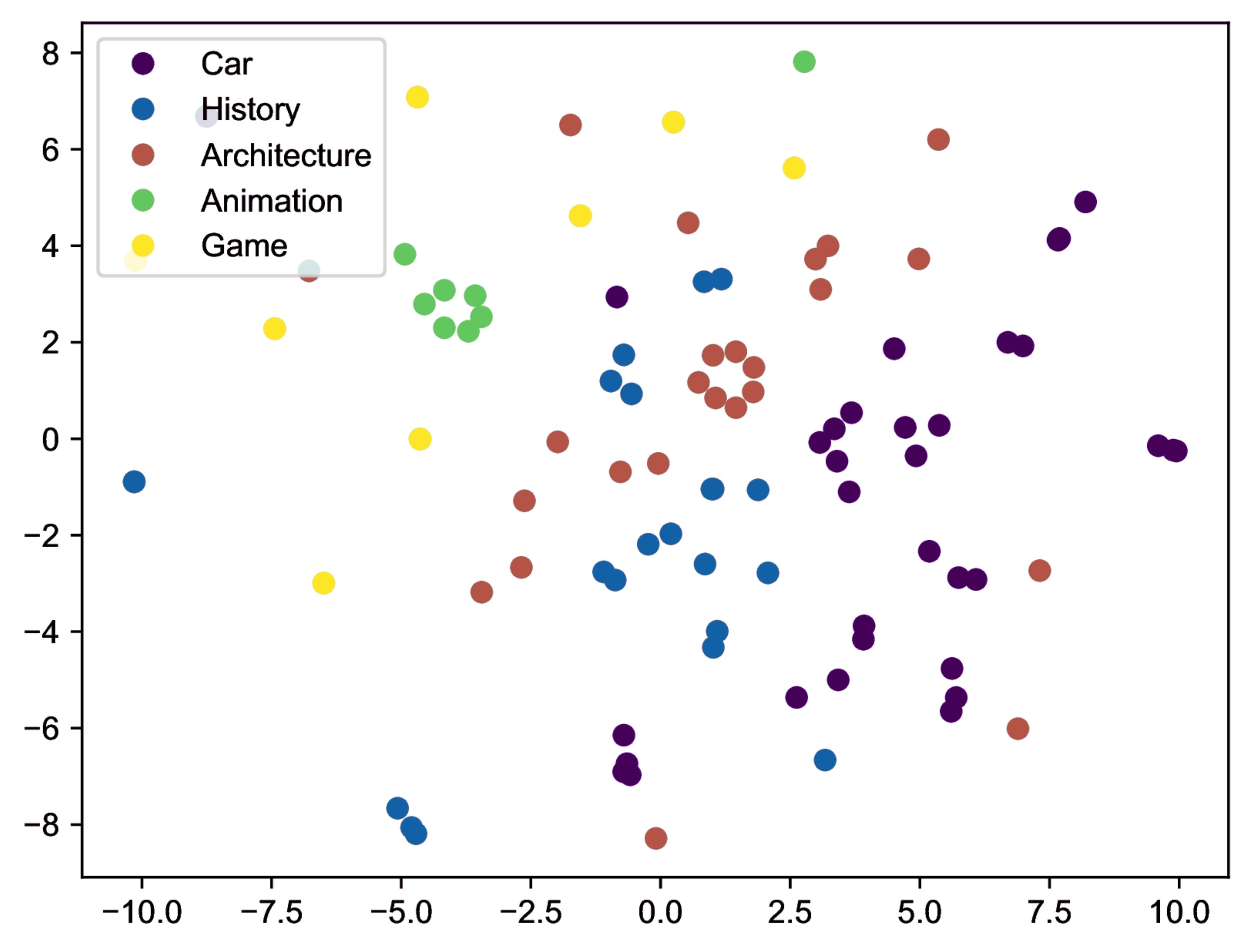}
\label{fig:ts1}}
\subfloat[Category \uppercase\expandafter{\romannumeral3}]
{\includegraphics[width=0.5\linewidth]{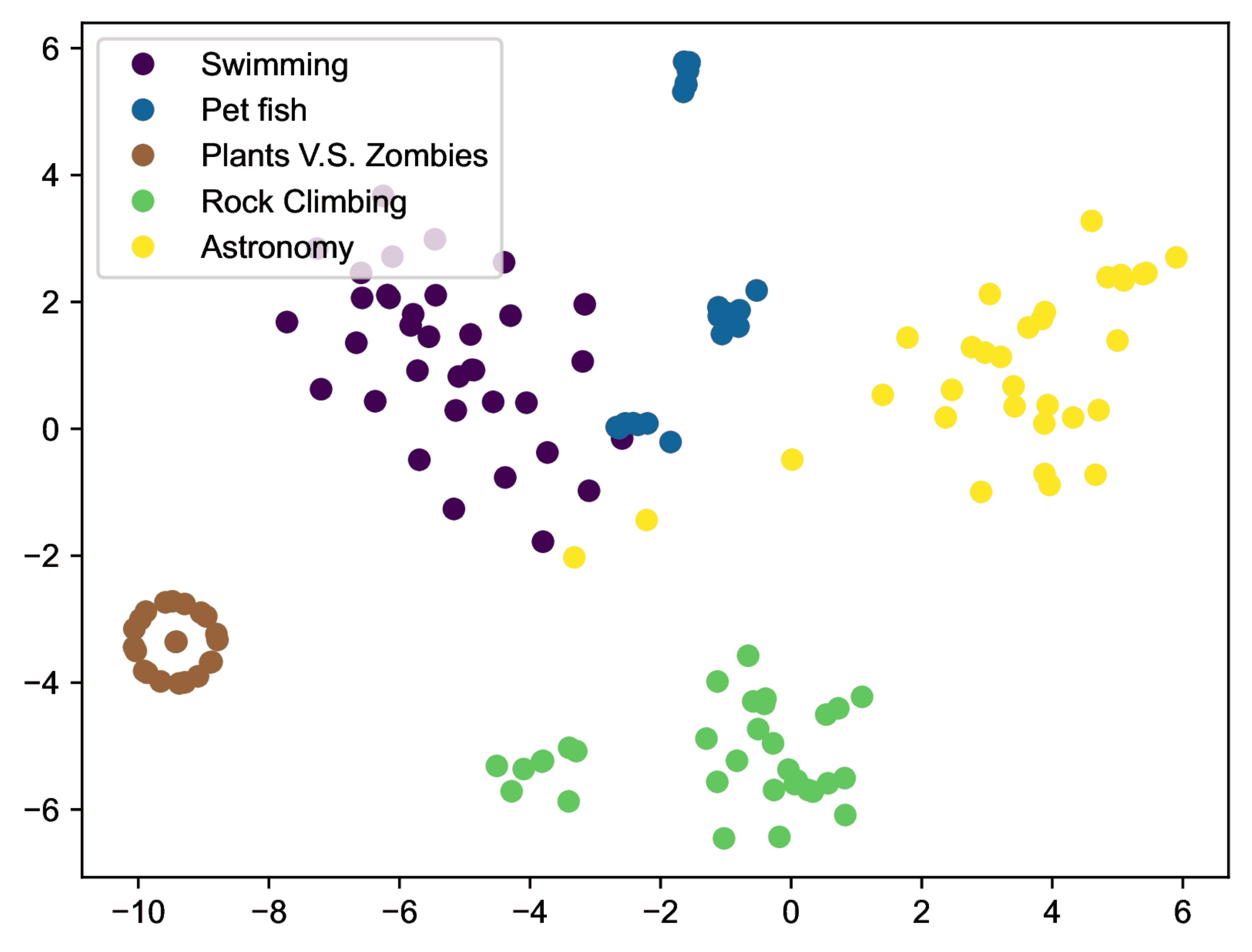}
\label{fig:ts3}}
\caption{Embedding visualization of videos with different (a) Category \uppercase\expandafter{\romannumeral1} and (b) Category \uppercase\expandafter{\romannumeral3} through t-SNE.}
\label{fig:ts}
\vspace{-3mm}
\end{figure}

Our dataset differs from existing ones by providing raw video files and preprocessed visual features. To assess the quality of these features, we use t-SNE for embedding visualization. We randomly select five video categories from both Category \uppercase\expandafter{\romannumeral1} and Category \uppercase\expandafter{\romannumeral3}. As shown in Figure~\ref{fig:ts}, videos from different categories exhibit distinct content features, confirming that the visual features effectively capture the semantic information of the videos. 



\subsection{Benchmarking Recommendation Algorithms}
\begin{table}[!t]
    \centering
    \caption{Benchmark results on 8 recommendation algorithms (including general and multimodal methods).}
    \label{tab:rec-results}
    \resizebox{\linewidth}{!}{
    \begin{tabular}{ccccc}
    \hline
        \textbf{Model} & \textbf{Recall@10} & \textbf{Recall@20}  & \textbf{NDCG@10} & \textbf{NDCG@20}  \\ \hline
        BPR & 0.0113 & 0.0218 & 0.0078 & 0.0114  \\ \hline
        LightGCN & 0.0223 & 0.0358  & 0.0169 & 0.0211  \\ \hline
        LayerGCN & 0.0208 & 0.0336  & 0.0160 & 0.0198  \\ \hline
        VBPR & 0.0184 & 0.0283  & 0.0140 & 0.0172  \\ \hline
        MMGCN & 0.0105 & 0.0187 & 0.0088 & 0.0114  \\ \hline
        GRCN & 0.0119 & 0.0224 & 0.0089 & 0.0125 \\ \hline
        LGMRec & 0.0159 & 0.0257 & 0.0131 & 0.0162 \\ \hline
        BM3 & \textbf{0.0238} & \textbf{0.0364} & \textbf{0.0178} & \textbf{0.0218}  \\ \hline
    \end{tabular}
    }
\vspace{-5mm}
\end{table}
Accurate user modeling and personalized recommendations are key focuses in both academia and industry. On short-video platforms, leveraging multimodal video information improves user modeling and recommendation quality. 
We evaluate 8 recommendation algorithms on our dataset in Table~\ref{tab:rec-results} to show the effectiveness of our dataset for user modeling. These algorithms are categorized into general and multimodal methods. The general methods include BPR~\citep{rendle2009bpr}, LightGCN~\citep{he2020lightgcn}, and LayerGCN~\citep{zhou2023layer}. The multimodal methods include VBPR~\citep{he2016vbpr}, MMGCN~\citep{wei2019mmgcn}, GRCN~\citep{wei2020graph}, BM3~\citep{zhou2023bootstrap}, and LGMRec~\citep{guo2024lgmrec}. 
We split the dataset into training, validation, and test sets with an 8:1:1 ratio, following the pipeline in~\citep{zhou2023comprehensive}. Results show that BM3 outperforms other methods by effectively utilizing multimodal data, consistent with results from other multimodal recommendation benchmarks~\citep{zhou2023comprehensive}. This validates the practical usage of our dataset for benchmarking recommendation algorithms.

\subsection{Filter Bubble Study} 
\begin{figure}[!t]
\centering
\subfloat[Active users.]{\includegraphics[width=0.51\linewidth]{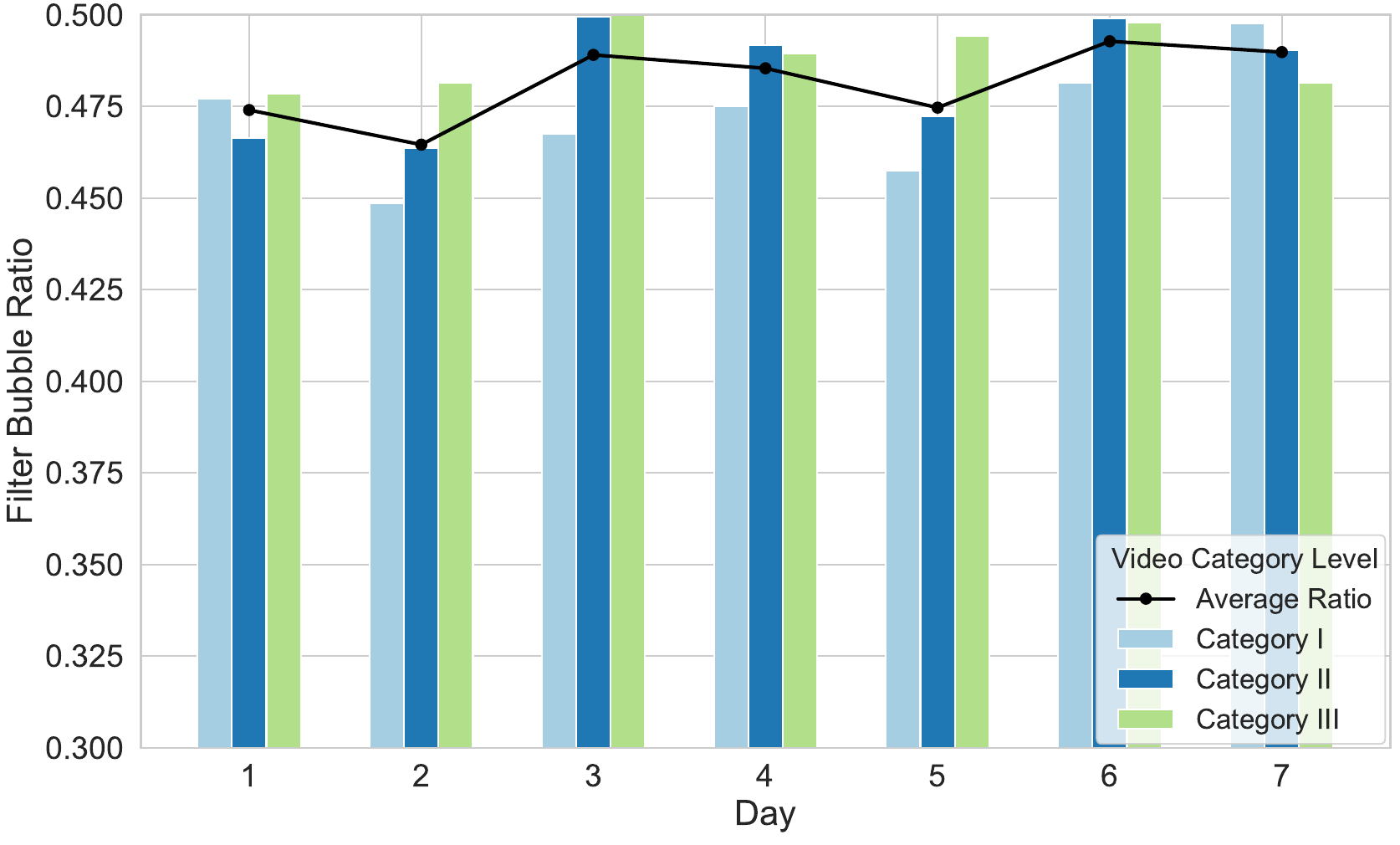}
}
\subfloat[Inactive users.]{\includegraphics[width=0.51\linewidth]{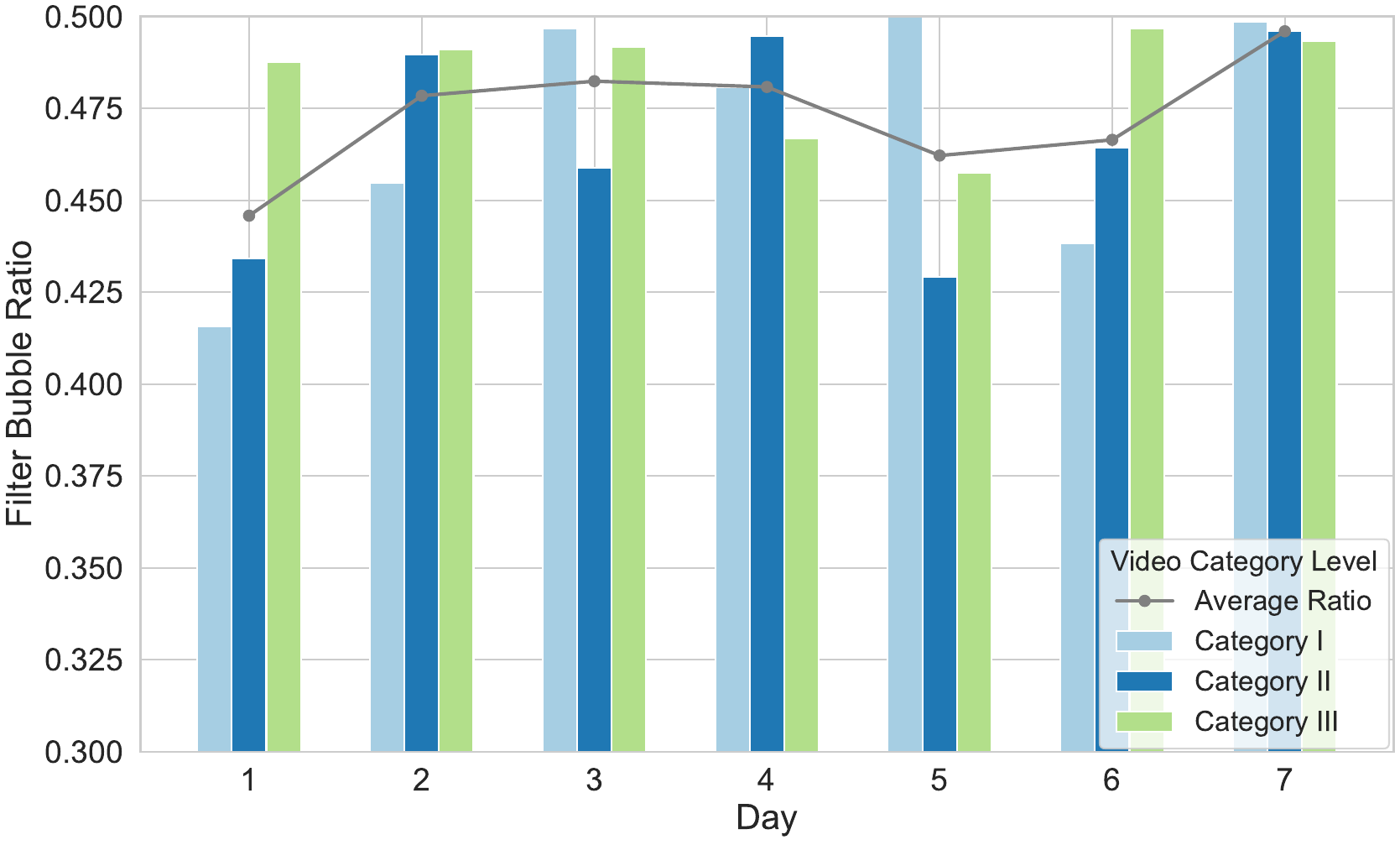}}
\caption{Analysis of the filter bubble ratio of active users (a) and inactive users (b) over time in our dataset.}
\vspace{-3mm}
\label{fig:filter bubble}
\end{figure}

The filter bubble phenomenon, where algorithms on online platforms reduce information diversity, has recently gained significant attention. Recommendation algorithms, which prioritize users' past preferences, tend to reinforce existing interests and limit exposure to diverse content, negatively impacting user experience and platform growth.
To measure content coverage, our dataset uses a 3-level video category system. Coverage for category \( c \) is defined as \( \frac{N_{seen}(u,c)}{N_{all}(c)} \), where \( N_{seen}(u,c) \) represents the number of categories a user \( u \) has interacted with, and \( N_{all}(c) \) is the total available categories at level \( c \). A user’s filter bubble extent is assessed by comparing this metric to the median value across all users. We track the evolution of filter bubble ratios over 7 days, distinguishing between active users ($\geq$3 videos/day) and inactive users (<3 videos/day). Results, shown in Figures~\ref{fig:filter bubble}, indicate that active users' filter bubble ratios remain stable with minor fluctuations, while inactive users' ratios steadily increase over time. This finding aligns with previous research~\citep{fu2024heavy}, suggesting that more engaged users do not necessarily have limited content preferences.

\section{Ethical Statement}
All participants have signed informed consent forms, agreeing to the use of their anonymized data for research. They have the right to opt out and request to remove their data at any time. User information has been anonymized to avoid user identification. The data collection was conducted strictly following the local ethical regulations similar to GDPR and was supervised by the Science and Technology Ethics Committee of Tsinghua University. 
The videos are collected from the public micro-video platform Kuaishou, we provide a collection of the existing public online videos with the ImageNet license. To avoid copyright issues, we will provide the original video URLs which can be accessed publicly. 
\section{Conclusion and Future Work}
In this paper, we introduce a large-scale dataset with rich user behavior, attributes and video content from a real mobile short-video platform, where we detail the data collection process, data characteristics and quality validation. Our dataset can support broad research communities such as user modeling, social science, human behavior understanding and so on. We have released the whole dataset and codes for data character analysis to facilitate relevant research. In the future, we plan to provide more fine-grained video content like semantic objects and sentiment labels so as to better describe video content. Moreover, we would like to process more data and provide user interactions with longer periods on the platform.


\bibliographystyle{ACM-Reference-Format}
\bibliography{sample-base}


\end{document}